\documentclass[prd,twocolumn,amsmath,amssymb,aps,nofootinbib]{revtex4-2}

\usepackage{epsfig}
\usepackage{url}
\usepackage{hyperref}

\usepackage[normalem]{ulem} 

\usepackage{latexsym}
\usepackage{epsfig}
\usepackage{amsmath}
\usepackage{amssymb}
\usepackage{wasysym}
\usepackage{graphicx}
\usepackage{dcolumn}
\usepackage{verbatim}
\usepackage[titletoc]{appendix}
\usepackage{amsfonts}
\usepackage{fancyvrb} %
\usepackage{tikz} %
\usetikzlibrary{calc}
\usepackage[export]{adjustbox}
\usepackage{capt-of} 

\graphicspath{{figures/}}

\def\cleancopy{1}  
\ifdefined\cleancopy
\newcommand{\new}[1]{#1}
\else
\newcommand{\new}[1]{{\color{blue}#1}}
\fi

\newcommand{\notebooklink}[1]{\href{#1}{\raisebox{-0.2ex}{\includegraphics[height=0.85em]{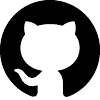}}}}

\usepackage{booktabs}

\usepackage[normalem]{ulem}%

\newcommand{\ba}{\begin{eqnarray}}
\newcommand{\ea}{\end{eqnarray}}
\newcommand{\be}{\begin{equation}}
\newcommand{\ee}{\end{equation}}

\definecolor{grey}{rgb}{0.4,0.4,0.4}
\definecolor{dullmagenta}{rgb}{0.4,0,0.4}
\definecolor{darkblue}{rgb}{0,0,0.4}
\definecolor{midblue}{rgb}{0,0,0.5}
\definecolor{midred}{rgb}{0.5,0,0}
\definecolor{orange}{rgb}{1,0.5,0}
\definecolor{lightbrown}{rgb}{0.75,0.5,0.25}
\definecolor{tan}{cmyk}{0.14,0.42,0.56,0}
\definecolor{djunglegreen}{cmyk}{0.99,0,0.52,0}
\definecolor{lightgreen}{rgb}{0,1,0}
\definecolor{olivegreen}{cmyk}{0.64,0,0.95,0.40}
\definecolor{midgreen}{rgb}{0.0,0.675,0.0}
\definecolor{darkgreen}{rgb}{0,0.5,0}
\definecolor{ceruleanblue}{rgb}{0.0, 0.2, 0.7}
\definecolor{burgundy}{rgb}{0.5, 0.0, 0.13}
\definecolor{hvred}{RGB}{186,12,47}

\hypersetup{
    colorlinks=true,
    linkcolor=ceruleanblue,
    filecolor=ceruleanblue,      
    urlcolor=midblue,
    citecolor=burgundy,
}

\newcommand{\mzc}[1]{}

\usepackage[all]{xy} %
\usepackage{amsfonts}

\makeatletter
\def\l@subsubsection#1#2{}
\makeatother

\begin{document}

\title{Lens Stochastic Diffraction: \\ A Signature of Compact Objects in Gravitational-Wave Data}

\author{Miguel Zumalac\'arregui}
\email{miguel.zumalacarregui@aei.mpg.de}
\affiliation{Max Planck Institute for Gravitational Physics (Albert Einstein Institute) \\
Am Mühlenberg 1, D-14476 Potsdam-Golm, Germany}

\begin{abstract}
Every signal propagating through the universe is diffracted by the gravitational field of intervening objects, known as gravitational lenses. 
Diffraction is most efficient when caused by compact objects, which form additional images of any source. 
Although secondary images are typically too weak to stand out in the noise, they can be detected collectively using gravitational waves (GWs) by leveraging 1) knowledge of the primary signal and 2) scaling of the signal's strength with the angular offset. 
The ensemble of secondary signals constitutes \textit{lens stochastic diffraction} (LSD): correlated Poisson-distributed fluctuations following a GW event due to intervening compact objects. 
The amplitude and temporal distribution of these signals encode \new{the abundance of the lenses, their mass spectrum and their density profiles}.
By including all secondary signals \new{over all resolved GW events, LSD offers an improvement of ${\sim}2.5$ orders of magnitude over the identification of individual lensed images,} for objects with mass $M_l\gtrsim 10^3 M_\odot$ and size $\lesssim 1{\rm pc}(M_l/10^4M_\odot)^{1/2}$.
\new{The framework generalizes to wave optics, where halos too diffuse to form images imprint localized features on the waveform.}
Developing data-analysis techniques will allow LSD to probe compact dark-matter halos and allow next-generation instruments to detect supermassive black holes, given the abundance expected from quasar luminosity studies.
\end{abstract}

\date{\today}

\maketitle

\paragraph*{\bf Introduction.}

Gravitational lensing, the deflection of propagating signals by intervening gravitational fields, is a sensitive probe of matter distribution. It illuminates the Universe's darkest objects: black holes~\cite{OGLE:2022gdj,Lam:2022vuq,Nightingale:2023ini,Wyrzykowski:2015ppa} and dark matter halos~\cite{Vegetti:2023mgp,Massey:2010hh,Clowe:2006eq,Vegetti:2012mc,Hezaveh:2016ltk,DiazRivero:2017xkd}. 
Most lensing applications to date rely on electromagnetic sources. 
However, the emergence of gravitational wave (GW) astronomy~\cite{LIGOScientific:2016aoc,LIGOScientific:2018mvr,KAGRA:2021vkt,Nitz:2021zwj,Olsen:2022pin,Mehta:2023zlk} provides new opportunities for gravitational lensing, which motivate detection strategies~\cite{Hannuksela:2019kle,McIsaac:2019use,Dai:2020tpj,Li:2023zdl,LIGOScientific:2023bwz,Janquart:2023mvf}. 
Given the steady \new{improvement} in sensitivity, multiply imaged GW sources are bound to become a reality in the near future~\cite{Dai:2016igl,Ng:2017yiu,Oguri:2018muv,Wierda:2021upe,Smith:2022vbp}. The detection of lensed GWs will provide new applications in cosmology, astrophysics and fundamental physics~\cite{Takahashi:2003ix,Xu:2021bfn,Jana:2022shb,Goyal:2020bkm,Goyal:2023uvm}, including the detection and characterization of dark-matter structures~\cite{Diego:2019rzc,Oguri:2020ldf,Oguri:2022zpn,Urrutia:2024pos,Ando:2026poq,Ando:2026eam,Guo:2022dre,Caliskan:2023zqm,Liu:2025ixi,Singh:2025uvp,Choi:2026lsa,Choi:2021bkx,Tambalo:2022wlm,Gais:2022xir,Christian:2018vsi,Basak:2021ten,Jung:2017flg,GilChoi:2023qrz,Fairbairn:2022xln,Urrutia:2021qak,Urrutia:2023mtk,Barsode:2024wda,Cheung:2024ugg,Savastano:2023spl,Liao:2018ofi,Jana:2024dhc,Seo:2023rjd,Vujeva:2025nwg,Zhou:2022yeo,Wang:2021lij}.

GW lensing is highly complementary to lensed electromagnetic radiation~\cite{Leung:2023lmq,Copi:2022ire,Tambalo:2022plm}. As GW detectors are sensitive to field amplitude and phase (rather than energy flux), the signal strength decreases with distance (rather than distance squared), facilitating the detection of remote sources. 
In lensing, this scaling makes GWs more robust to demagnification, allowing the observation of faint images, including strongly deflected trajectories for sources near a massive black hole~\cite{Kocsis:2012ut,Gondan:2021fpr,Oancea:2022szu,Leong:2024nnx,Ubach:2025dob,Oancea:2023hgu}, central images of strongly lensed systems \cite{Tambalo:2022wlm,Hezaveh:2015oya}, and systems with large source-lens angular separation \cite{Takahashi:2003ix,Gao:2021sxw,Caliskan:2022hbu,Savastano:2022jjv,Savastano:2023spl,Caliskan:2023zqm}. The prospects of observing faint images \new{increase} significantly because the number of lenses scales with the square of source-lens separation. 

In addition to loud transient signals, detectors are also sensitive to GW backgrounds, a stochastic superposition of many faint signals. Astrophysical backgrounds from unresolved mergers are expected~\cite{Regimbau:2022mdu,Lehoucq:2023zlt,Babak:2023lro,Rieck:2023pej}, and cosmological backgrounds could be produced by early-universe phenomena~\cite{Caprini:2018mtu}. Although signals are too weak to be resolved, their existence can be inferred as an excess over instrumental noise or by cross-correlating the output of multiple detectors~\cite{Christensen:2018iqi,Renzini:2022alw,vanRemortel:2022fkb}. 
Evidence for a GW background at nanohertz frequencies was recently reported~\cite{EPTA:2023xxk,NANOGrav:2023hvm,Figueroa:2023zhu}.

Here, I present \textit{lens stochastic diffraction} (LSD), a novel signature of compact structures consisting of faint counterparts present after loud GW events. These secondary signals are caused by compact gravitational lenses: Their waveform is fixed by the main event, but their distribution in time is stochastic, resembling that of GW backgrounds.
I will introduce the LSD signal caused by an ensemble of point lenses, derive its time distribution and signal-to-noise ratio, and show its potential to detect compact objects, before discussing open issues and prospects. \new{The analysis is restricted to the geometric-optics regime, in which lensing produces discrete images of the source; the generalization to wave optics is presented in the End Matter, Appendix~E. Technical details are provided in the remaining appendices, and all results can be reproduced with publicly available notebooks.}\footnote{\new{\url{https://github.com/miguelzuma/lsd-gw-notebooks}}} Units assume $\new{\hbar} = c=1$.

\vspace{5pt}
\paragraph*{\bf Lens stochastic diffraction.}

\new{In the geometric-optics regime, lensed signals are fully described by discrete images: the stationary points $\vec x_I$ of the Fermat potential, with magnifications $\mu_I^{-1} = \det(\partial \vec y_l/\partial \vec x)|_{\vec x_I}$~\cite{Schneider:1992}.}
A point-like gravitational lens forms at least one additional image of any source. The ratio of the lensed/unlensed flux is given by the magnification, 
which for an isolated point lens reads
\begin{equation}\label{eq:mu_pl}
    \mu_I =  \frac{\Delta_I^4}{\Delta_I^4-1} = \frac{1}{2}\pm \frac{y_l^2+2}{2y_l\sqrt{y_l^2+4}}\,.
\end{equation}
Here $I=(+,-)$ labels the main/additional image and the sign corresponds to the parity. $\Delta_{I} = |\vec x_I - \vec y_l|$ is the image-lens angular separation: coordinates are centered around the undeflected source and $\vec x_I\equiv \vec \theta_I/\theta_E$, $\vec y_l\equiv \vec \theta_l/\theta_E$ are the image and lens position in units of the Einstein angle, $\theta_E = \sqrt{\frac{4G M D_{\rm LS}}{D_{\rm L} D_{\rm S}}}$. The angular diameter distances $D_{\rm L}$, $D_{\rm S}$, $D_{\rm LS}$ (for observer--lens, observer--source and lens--source, respectively) are given by flat-$\Lambda$CDM cosmology~\cite{Planck:2018vyg}. 

The existence of a faint image near the lens is associated with a strongly deflected trajectory.
At large angular separations, $y_l \gg 1$, the secondary image flux scales as $|\mu_{-}| \approx y_l^{-4}$. %
In typical electromagnetic sources, strong demagnification prevents the observation of additional images in generic source-lens configurations. 
Instead, the signal amplitude of GWs is modulated by $\sqrt{|\mu_{-}|}\approx y_l^{-2}$, making faint images more easily observable.

Increasing the offset $y_l$ decreases the amplitude of individual images but increases their number: N point lenses produce at least $N+1$ images and as many as $(N+1)^2$~\cite{Witt_point_lenses_90}.%
\footnote{Even more images form when lenses are distributed at different distances, because of consecutive deflections.
I will neglect those and assume that all lenses can be projected onto a common lens plane by rescaling the Einstein radii~\cite{CaganSengul:2020nat}.}
The enclosed average number of lenses
\begin{equation}\label{eq:enclosed_N}
    N_c(<y_l) = \kappa_c\, y_l^2 \,,
\end{equation}
depends on the line of sight \new{$\hat n$} via the convergence $\kappa_c = 4\pi G {\Sigma_c} \frac{D_{\rm LS}D_{\rm L}}{D_{\rm S}}$, where ${\Sigma_c}$ \new{is} the projected surface density of compact objects\new{,
\begin{equation}\label{eq:convergence_los}
\kappa_c(z_S,\hat n) = 4\pi G\int_0^{z_S}\frac{dz'}{(1+z')H(z')}\frac{D_{\rm L} D_{\rm LS}}{D_{\rm S}}\,\rho_c(z',\hat n)\,,
\end{equation}
where $\rho_c(z',\hat n)$ is the physical density of compact objects along the line of sight and $H(z)$ the expansion rate. Because each lens enters only through its angular offset in units of its Einstein angle $\theta_E(M_l,z',z_S)$, all objects between observer and source are projected onto a single effective convergence, independent of the lens mass}. 
The lens number $K$ follows a Poisson distribution $\mathcal{P}(K|N_c)$, with $\kappa_c$ set by the line of sight to the source. The average convergence is~\cite{Schneider:1992} %
\begin{equation}\label{eq:aconvergence_avg}
\bar \kappa_c(z_S) = \frac{3}{2}f_c\Omega_{\rm DM} H_0^2\int_0^{z_S}dz'\frac{(1+z')^2}{H(z')}\frac{D_{\rm L} D_{\rm LS}}{D_{\rm S}}\,, 
\end{equation}
where $H_0$ is the Hubble constant and $f_c$ is the fraction of compact objects, relative to the dark-matter abundance $\Omega_{\rm DM}$\new{: Eq.~\eqref{eq:aconvergence_avg} follows from Eq.~\eqref{eq:convergence_los} with $\rho_c = f_c\Omega_{\rm DM}\rho_{\rm cr,0}(1+z')^3$, i.e.}~the comoving density of compact objects is assumed constant.
\new{Dark-matter clustering induces line-of-sight variance in Eq.~\eqref{eq:convergence_los}: signals traversing dense regions receive stronger LSD. Because lens contributions are additive at low optical depth, averages over the source population (Eq.~\ref{eq:rho_LSD} below) are unaffected --- and conservative, as high-$\kappa_c$ lines of sight are further boosted by additional images and collective lensing effects (Fig.~\ref{fig:N0_SNR}).} 
\begin{figure}[t]
    \centering
    \includegraphics[width=\columnwidth]{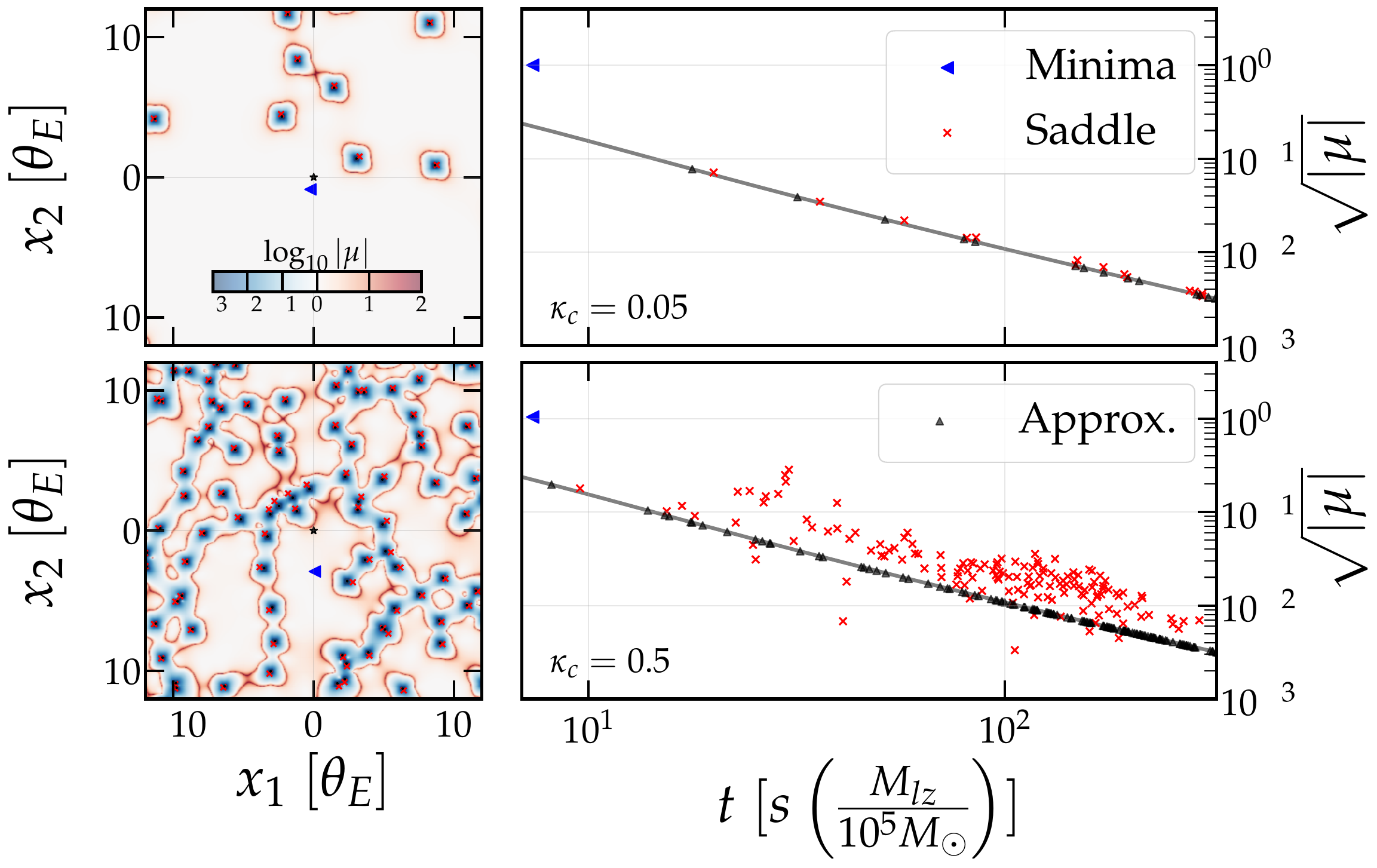}
    \caption{Distribution of images for low (top\new{, $\kappa_c=0.05$}) and high (bottom\new{, $\kappa_c=0.5$}) projected lens density.
    \textbf{Left:} lens (dots) and images (triangles, crosses), color shows \new{the local magnification (red/blue for high/low, colorbar)}.
    \textbf{Right:} amplitude and arrival time of the secondary images (crosses) \new{and} the isolated-lens approximation (dots).
    The primary image (triangle) is shifted from $t=0$. %
    \notebooklink{https://nbviewer.org/github/miguelzuma/lsd-gw-notebooks/blob/main/LSD_realizations.ipynb\#Image-realizations}
    }
    \label{fig:image_realizations}
\end{figure}

For point lenses $\bar\kappa_c$ is independent of the lens' mass. %

Scaling with $y_l$ suggests that secondary images will rarely be identified individually but can be detected collectively.
Figure~\ref{fig:image_realizations} shows the distributions of images in the sky and their amplitude and delay, for two realizations with low and high $\kappa_c$\new{, computed exactly from the joint Fermat potential of all lenses (End Matter, Appendix A)}.
For $\kappa_c\ll 1$ at most one lens is closely aligned with the source. Then Eq.~\eqref{eq:mu_pl} describes the two brightest images, plus other secondary images with $y_l\gg 1$.
The higher $\kappa_c$ increases both individual magnifications and the chance of forming additional images~\cite{Katz_random_microlensing_86,Venumadhav:2017pps,Pascale:2021bdo}.
Therefore, the isolated lens approximation is accurate for $\kappa_c\ll 1$ and conservative otherwise \new{(End Matter, Appendix A, Fig.~\ref{fig:N0_SNR})}.
Hereafter I assume that lenses can be treated as isolated.

The time distribution of secondary images encodes information about the lens mass distribution.
Each secondary image arrives with a characteristic delay, which depends on the lens mass, redshift and offset  as
\begin{equation}
\label{eq:time_delay}
t_l \approx 2GM_l (1+z_l) y_l^2 \equiv \tilde t_l \frac{y_l^2}{2} \quad  (y_l\gg 1) \,,
\end{equation}
where the second equality defines a dimensionless delay.
To remain in the geometric-optics description, we will require that the time delays between different images satisfy $2\pi (t_i- t_j)f\gg 1$, where $f$ is the GW frequency~\cite{Tambalo:2022plm}.

A GW signal lensed by a collection of point-particles in the geometric-optics regime is described by
\begin{equation}\label{eq:lensed_time_domain}
    h_L(t) = \sqrt{\mu_+}h_0(t) + \sum_I\sqrt{|\mu_{I}|}h_0^{\dagger}(t-t_I) \,, %
\end{equation}
Here $\mu_+>1$ is the primary image magnification, not directly observable, and $h_0(t)=\sum_{p=+,\times} F_p(t)h_{0,p}(t)$ includes the detector antenna pattern.
The index $I$ labels the additional images and $h_0^\dagger$ is the Hilbert-transform of the unlensed signal, accounting for the phase difference in negative parity images \cite{Dai:2017huk,Ezquiaga:2020gdt}.%
\footnote{
Positive parity images can be included in Eq.~\eqref{eq:lensed_time_domain} by adding $\sum_M\sqrt{\mu_M}h_0(t-t_M)-\sum_N\sqrt{\mu_N}h_0(t-t_N)$, where $M,N$ index secondary minima and maxima of the time delay, respectively.
}
Under the sparsity assumption there is one image per lens $I\to l$ and $\mu_{I}\approx \mu_{-}(y_l)$ is well approximated by Eq.~\eqref{eq:mu_pl}.

The \textit{lens stochastic diffraction} (LSD) signal is defined by the sum of additional images in Eq.~\eqref{eq:lensed_time_domain}.
Although the time distribution is stochastic, LSD differs from GW backgrounds:
\begin{enumerate}
    \item  The waveform of individual LSD contributions is known from the primary signal $h_0$ (Eq.~\ref{eq:lensed_time_domain}), up to parameter-estimation uncertainties.
    \item LSD is strongly direction dependent, with the same sky localization as the main signal. GW backgrounds are approximately isotropic (extragalactic/cosmological) or correlated with the Milky Way. 
    \item LSD is neither stationary nor Gaussian: its correlation with the primary GW signals is determined by the lens mass distribution (Eq.~\ref{eq:time_delay}). 
\end{enumerate}
LSD also differs from the contribution of magnified high redshift sources to the astrophysical GW background~\cite{Buscicchio:2020cij,Buscicchio:2020bdq,Mukherjee:2020tvr}.

\begin{figure}[t]
    \centering
\includegraphics[width=1.\columnwidth]{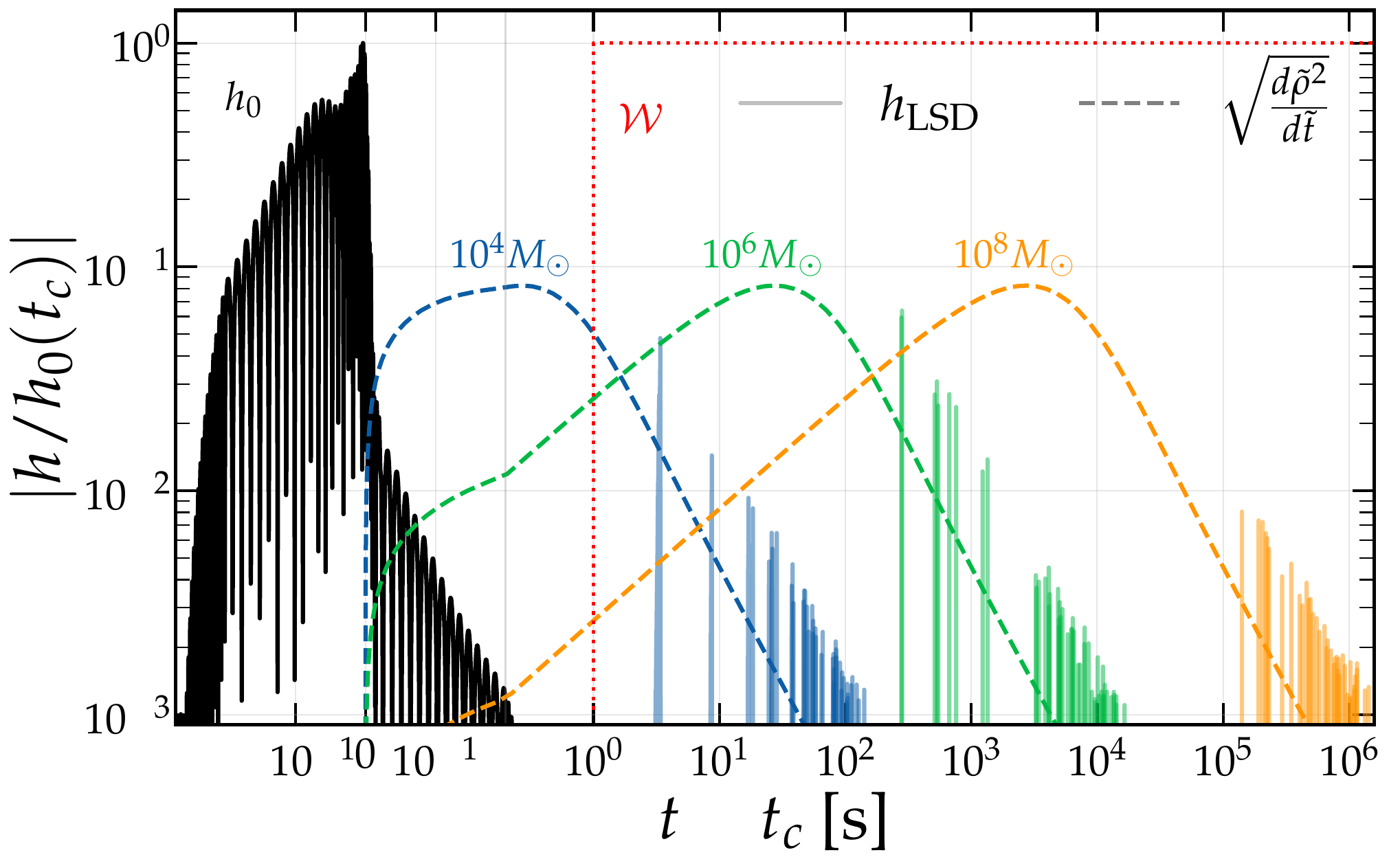} \quad
    \caption{
    Lens stochastic diffraction. Solid lines show the strain corresponding to the main image (black) and 3 different LSD realizations for \new{different} lens masses, assuming $\bar \kappa_c = 0.05$. 
    Dashed lines show the average LSD amplitude, dotted line shows the window used to define the LSD signal. 
    \notebooklink{https://nbviewer.org/github/miguelzuma/lsd-gw-notebooks/blob/main/LSD_exploration.ipynb\#LSD-in-the-time-domain}
    }
    \label{fig:lsd_pedagogical}
\end{figure}

Examples of LSD realizations following a $30+30M_\odot$ equal-mass, non-spinning merger are shown in Fig.~\ref{fig:lsd_pedagogical} for different lens masses. 
The projected density $\bar \kappa_c=0.05$ corresponds to a source at $z_S=1$ with $f_c=1$.

\vspace{5pt}
\paragraph*{\bf Temporal distribution and signal-to-noise ratio.}

Let us now investigate the temporal distribution and signal-to-noise ratio (SNR) of LSD. To separate LSD in Eq.~\eqref{eq:lensed_time_domain}, I will introduce a window function $\mathcal{W}(t)$ that is one for $t-t_c>T_{\rm min}=1$ s after coalescence and zero elsewhere (Fig.~\ref{fig:lsd_pedagogical}). The LSD signal is $\Delta h(t)\equiv \mathcal{W}(t) h_L(t)$.
\new{The choice $T_{\rm min}=1$\,s is conservative: it removes any overlap between the primary signal's ringdown and the secondary images without further waveform modelling. The results below are insensitive to this choice: the window only suppresses signals with characteristic delays $\tilde t_l\lesssim T_{\rm min}$, so varying $T_{\rm min}\in[0.1,10]$\,s shifts the low-mass edge of the sensitivity in proportion to $T_{\rm min}$, leaving the high-mass limits unchanged (End Matter, Appendix B).}

The matched-filter SNR can be used because the waveform of secondary signals is known. Using $\rho^2_{\rm LSD} = (\Delta h|\Delta h)$, where $(h_1|h_2)$ is the noise-weighted inner product~\cite{Lindblom:2008cm}, %
the \textit{LSD-SNR} is
\begin{equation}\label{eq:lsd_snr}
\frac{\rho^2_{\rm LSD}}{\rho_0^2}  = \sum_I |\mu_{I}| \new{\mathcal{W}(t_I)} + \sum_{I\neq J}\sqrt{|\mu_I\mu_J|} M\left(t_I-t_J\right)\new{\mathcal{W}(t_I)\mathcal{W}(t_J)}\,,
\end{equation}
where $\rho_0^2\equiv (h_0^\dagger|h_0^\dagger)=(h_0|h_0)$ is the SNR of the unlensed event and $M(\Delta t) = \rho_0^{-2}\left(h_0(t)|h_0(t-\Delta t)\right)$ is the match between different images.
The first sum in Eq. \eqref{eq:lsd_snr} represents the contribution of each image \new{arriving within the window, $t_I > T_{\rm min}$}.
The double sum accounts for interference between separate images and has zero average, since $t_I$'s are not correlated.
Eq.~\eqref{eq:lsd_snr} assumes a constant antenna pattern, a good approximation for $t_I\ll 1$d, or $M_l\lesssim 10^7 M_\odot$.%
\footnote{For large $t_I,M_l$ the average LSD-SNR is slightly lower due to selection bias, as a detection threshold favors events with $F_p(t_c)>\langle F_p\rangle$.} %

Let us now study the average LSD-SNR relative to the unlensed event, $\tilde\rho^2_{\rm LSD} \equiv \left\langle {\rho^2_{\rm LSD}}\right\rangle/\rho_0^{2}$. The sum over images is replaced by integrals weighted by the lens number density $\sum_I\to \int dt \frac{dN_c}{dt} \mathcal{W}(t)$.
Assuming isolated lenses, the average \textit{relative differential LSD-SNR} is
\begin{equation}\label{eq:lsd_snr_dt}
    \frac{d\tilde\rho^2_{\rm LSD}}{dt dM} = \mathcal{W}(t)\frac{df_c}{d M}\int_0^{z_S}{dz'} dy^2\frac{d\bar \kappa_c}{dz'}|\mu_{-}|(y)\delta(t-t(y))\,,
\end{equation}
Here $\frac{d\bar \kappa_c}{dz'}$ is the derivative of Eq.~\eqref{eq:aconvergence_avg}, $\frac{df_c}{dM}$ is the lens mass function and the average lens density has been used (Eq.~\ref{eq:enclosed_N}). The magnification and time delay are evaluated as for an isolated lens, cf.~Eqs.~(\ref{eq:mu_pl},\ref{eq:time_delay}).
\new{Eq.~\eqref{eq:lsd_snr_dt} is linear in the lens mass function $df_c/dM$: an extended mass spectrum contributes as the superposition of monochromatic populations (the colored lines of Fig.~\ref{fig:lsd_pedagogical}), each entering at its characteristic time delay (Eq.~\ref{eq:time_delay}), as shown explicitly in the End Matter, Appendix~B (Fig.~\ref{fig:massfunction}).}

The total LSD-SNR is the integral of Eq.~\eqref{eq:lsd_snr_dt} over time and lens mass. If $\mathcal{W}(t)\to 1$ it simplifies to
\begin{equation}\label{eq:lsd_rho2_avg}
    \tilde\rho^2_{\rm LSD} = 2 \bar \kappa_c \int dy_l \,y_l \, |\mu_{-}|(y_l) \approx \bar \kappa _c\,.
\end{equation}
Note that including $\mathcal{W}(t)$ will add a dependence on both the lens' mass and the source redshift, which is important for $M\lesssim 10^4M_\odot$ (cf.~Fig.~\ref{fig:lsd_pedagogical}). 
Similarly, an upper limit on $\mathcal{W}(t)$ will affect the sensitivity to high mass objects. %

The distribution of the LSD-SNR depends strongly on $\kappa_c$, whose summary statistics are shown in Fig.~\ref{fig:N0_SNR} \new{(End Matter, Appendix A)}.
For the isolated lens approximation, the mean of random realizations agrees with Eq.~\eqref{eq:lsd_rho2_avg}.
At low $\kappa_c$, the distribution is skewed, as the median $\propto \kappa_c^2\ll\tilde\rho^2$ deviates from the mean due to rare events with high $\tilde\rho_{\rm LSD}$ ($y_l\ll 1$). %
The simulated LSD-SNR becomes larger than the approximate value \new{for} $\kappa_c\gtrsim 0.1$ due to the formation of additional images and collective lensing effects, confirming that the isolated lens approximation is conservative.

\vspace{5pt}
\paragraph*{\bf Observational prospects.}

Let us now estimate the sensitivity of ground detectors to LSD given the abundance of compact objects, $f_c$. The rate calculation follows Refs. \cite{Dominik:2014yma,Chen:2017wpg}. I consider the LIGO-Virgo-KAGRA (LVK) detector network at O5 sensitivity, the Einstein Telescope (ET)~\cite{Maggiore:2019uih} and a single Cosmic Explorer (CE) interferometer~\cite{Evans:2021gyd}. I will consider only non-spinning and quasicircular black hole mergers.

The detection rate and the average LSD-SNR \new{accumulated per unit observation time over the} full catalogue of sources are
\begin{align}
\dot N &= \mathcal{R}_0 \int d\vec \theta \frac{dP}{d\vec \theta} \int dz \frac{dV}{dz} P_{\text{det}} f_\text{merge} \,, \label{eq:rate}
\\
\new{\rho_{\rm LSD,pop}^2} &= \mathcal{R}_0\int d\vec \theta \frac{dP}{d\vec \theta} \int dz\, \bar \kappa_c \frac{dV}{dz} \bar{w}^2 \rho_{\text{opt}}^2 P_{\text{det}} f_\text{merge} \,.\label{eq:rho_LSD}
\end{align}
Here $\vec \theta=\{m_1,m_2\}$ are the binary intrinsic parameters and $ \frac{dP}{d\vec \theta}$ is the population model given by the ``power law + peak'' model (fiducial parameters in Ref.~\cite{Talbot:2018cva}). 
The volume element $\frac{dV}{dz}$ is that of flat $\Lambda$CDM~\cite{Planck:2018vyg}\new{, $f_\text{merge}(z)$ is the redshift evolution of the merger rate, normalized to $f_\text{merge}(0)=1$,} and $\mathcal{R}_0 = 30 {\rm Gpc}^{-3}{\rm yr}^{-1}$ is the local merger rate \cite{KAGRA:2021duu}.
\new{I adopt a constant comoving rate, $f_\text{merge}=1$: current population analyses constrain the low-redshift slope of the merger rate but not its high-redshift evolution~\cite{KAGRA:2021duu}, and among the standard choices this one gives the weakest limits. A star-formation-rate evolution~\cite{Madau:2014bja} would strengthen them by a factor ${\sim}4$ for every network, and a pure power law $(1+z)^{2.7}$, which never turns over, by a factor ${\sim}400$ for ET and CE (End Matter, Appendix B, Table~\ref{tab:rates}).}
$P_{\rm det}$ is the fraction of detected sources: writing the observed SNR as $\rho_{\rm obs}= w\,\rho_{\rm opt}$, with $w\in[0,1]$ the antenna-pattern projection factor, $P_{\rm det} = P\left(w>\rho_{\rm th}/\rho_{\rm opt}(z,\vec\theta)\right)$ with $\rho_{\rm th}=8$. 
$P_{\rm det}$ and $\rho_{\rm opt}$ are obtained using \texttt{gwfast} \cite{Iacovelli:2022bbs,Iacovelli:2022mbg}. 
The LSD-SNR depends on the second moment of the projection factor, $\bar w^2 = \int dw\,\frac{dP}{dw} w^2$, and on $\bar \kappa_c(z)\propto f_c$, Eq.~\eqref{eq:aconvergence_avg}.
\new{The subscript distinguishes the population-accumulated LSD power per unit observation time from the per-event quantities of the previous section, $\tilde\rho^2_{\rm LSD}$.}

The mass of the lens affects the total LSD-SNR by changing the arrival time of secondary signals (Eq.~\ref{eq:time_delay}) relative to $T_{\rm min}$ in the window function.
Eq.~\eqref{eq:rho_LSD} neglects the effect of windowing, i.e. $4GM_l(1+z_l)\gg T_{\rm min}$: The effect of $M_l$ is estimated by multiplying $\new{\rho^2_{\rm LSD,pop}}$ by $\frac{1}{\tilde\rho^{2}}\int_{T_{\rm min}}^\infty\frac{d\tilde\rho^2}{dt}dt$ (Eq.~\ref{eq:lsd_snr_dt}), assuming $z_S=1$ (below the typical source redshift for future detectors).
Note that LSD includes lensed events where secondary signals are individually detected, $\rho_0\sqrt{|\mu_-|}>\rho_{\rm th}$.

\new{Sensitivity forecasts follow from Poisson statistics: the number of secondary signals, and hence the accumulated LSD power, scales linearly with $f_c$ through $\bar\kappa_c$. A null result over an observation time $T_{\rm obs}$ excludes at 90\% confidence level (c.l.) abundances $f_c > 2.3/[\rho^2_{\rm LSD,pop}(f_c{=}1)\,T_{\rm obs}]$, where $2.3$ is the 90\% Poisson factor for zero counts. Details of the computation, including the window degradation at finite $M_l$, are given in the End Matter, Appendix B.}

\begin{figure}[t!]
    \centering
    \includegraphics[width=\linewidth]{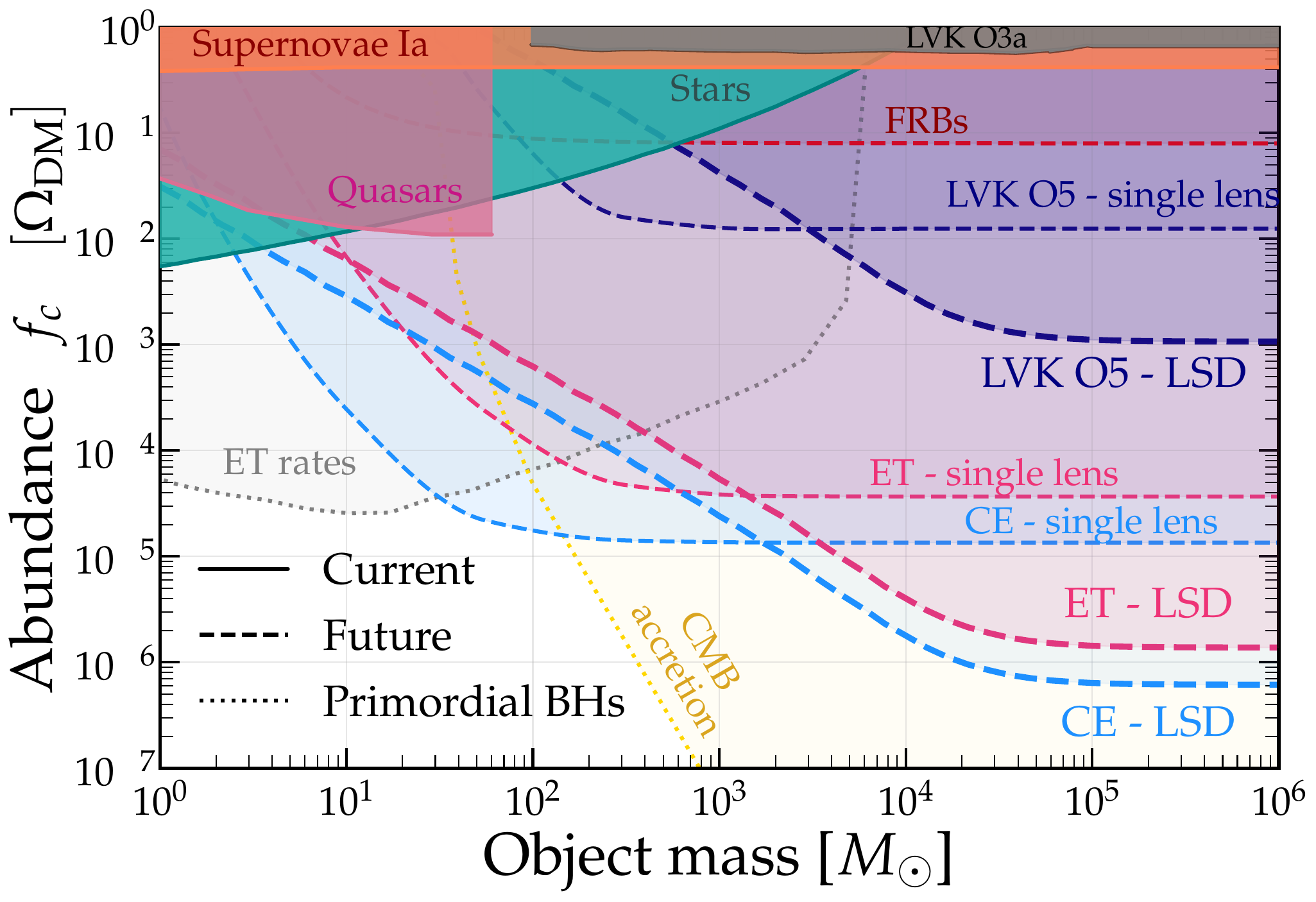}
    \caption{Potential of LSD to constrain the abundance of compact objects (thick dashed) for LVK-O5 (blue), ET (magenta) and CE (light blue) after 1 yr. For high masses, LSD results in a $\sim 10^2-10^5$ improvement over existing lensing limits (solid). The gain is also substantial relative to \new{standard single-lens GW analyses} by the same networks (thin dashed\new{, labeled ``single lens''}).
    Model-dependent results for primordial black holes (dotted) are shown for comparison, see text.
    \notebooklink{https://nbviewer.org/github/miguelzuma/lsd-gw-notebooks/blob/main/LSD_forecast.ipynb\#Sensitivity-forecasts}
    }
    \label{fig:forecast}
\end{figure}

Detection of LSD or its absence provides competitive limits on the abundance of compact objects with $M\gtrsim 10^3 M_\odot$.
Fig.~\ref{fig:forecast} shows current lensing constraints from quasars~\cite{Esteban-Gutierrez:2023qcz}, stars~\cite{Mroz:2024mse,Blaineau:2022nhy,Oguri:2017ock}, type Ia supernovae~\cite{Zumalacarregui:2017qqd} and GWs from LVK O3a \cite{Basak:2021ten}, as well as expected limits from fast radio bursts (FRBs)~\cite{Munoz:2016tmg}, GWs under single-lens analysis~\cite{Jung:2017flg,GilChoi:2023qrz} and with LSD (90\% c.l. and 1 year of observation). 
Lensing limits for $M\gtrsim 300 M_\odot$ will be dominated by lensed GWs in the near future. 
Single-lens \new{analyses} are more stringent for objects with $M\sim (8\pi G f)^{-1}$ due to wave-optics lensing effects (excluded from LSD by signal windowing). Instead, for $M\gg (8\pi G f)^{-1}$, LSD incorporates information from low-amplitude images\new{. Counting only the brightest secondary image above a search threshold $\rho_{\rm img}=6$ (an optimistic benchmark for sub-threshold searches of additional images) under identical population assumptions and time window yields limits weaker than LSD's by ${\sim}2.5$ orders of magnitude for all networks (End Matter, Appendix C). The counting limits flatten at high mass like the dedicated single-lens forecasts~\cite{Jung:2017flg,GilChoi:2023qrz} and the LVK O3 search~\cite{Basak:2021ten}, validating the comparison.} %

LSD constraints are complementary to other probes of compact objects and dark matter. For comparison, Fig.~\ref{fig:forecast} shows two model-dependent limits for primordial black holes: current cosmic microwave background (CMB) accretion limits~\cite{Serpico:2020ehh} and forecasted limits from merger rate by ET~\cite{Kalogera:2021bya}. %
Although stringent, these limits require dark matter to be very compact and be present before recombination. In contrast, LSD probes extended structures, such as compact dark-matter halos and black holes of astrophysical origin. Then, the finite size prevents the formation of additional images for offsets $y \gtrsim \Delta x^{-1}$, where $\Delta x = \frac{R_L}{D_L\theta_E}$ is the object's size over its Einstein radius. This allows objects with
$R_L\lesssim  1{\rm pc}(M_l/10^4M_\odot)^{1/2}$ to be probed by LSD, $\sim 10^3$ times larger than the CMB~\cite{Croon:2024rmw}.
Such a cutoff for secondary images in $y_l$ causes a drop in the LSD after a characteristic delay (Eq.~\ref{eq:time_delay}), whose observation would reveal the lens' typical size.
Non-singular lenses also form central images~\cite{Tambalo:2022wlm}, whose short time delay may introduce wave-optics distortions. The analysis here can be extended by replacing $h_0^\dagger$ with the distorted waveform from the extended lens.
The effect of the lens mass profile can be determined accurately by changing $\mu(y)$ in Eq.~\eqref{eq:lsd_snr_dt}\new{: examples for homogeneous spheres, cuspy profiles and dark-matter halos are given in the End Matter, Appendix~D (Fig.~\ref{fig:extended})}.%

Two potential targets of LSD are supermassive black holes (SMBHs) and compact dark-matter halos. Both ET and CE have the potential to probe SMBHs via LSD, with $\Omega_{\rm SMBHs}/\Omega_{\rm DM}\sim 1.5\cdot 10^{-5}$ estimated from the quasar luminosity function~\cite{Hopkins:2006fq,Soltan:1982vf}. 
Other systems, such as lensed gamma-ray bursts, suggest the existence of compact lenses whose mass and abundance could be further probed by LSD~\cite{Paynter:2021wmb,Yang:2021wwd,Kalantari:2021sqy}.
Beyond detection, the temporal distribution of the LSD (Eq.~\ref{eq:lsd_snr_dt}) can be used to constrain the mass function of compact structures.

\vspace{5pt}
\paragraph*{\bf Discussion.}

Lens stochastic diffraction (LSD) is a new signature in GW data sensitive to compact objects with $M_l\gg (Gf)^{-1}$. It consists of secondary signals after a GW event, each a low-amplitude, phase-shifted copy of the primary signal (Figs.~\ref{fig:image_realizations}~\&~\ref{fig:lsd_pedagogical}). The time distribution follows Poisson statistics, depending on the abundance of lenses and their mass. 
Secondary signals can be identified individually only if a lens is closely aligned with the source. 
Instead, LSD considers them collectively, thus including information from lenses at large angular separations, too weak for individual identification.
LSD, with deterministic waveform but stochastic time distribution, is distinct from both individual events and GW backgrounds.

LSD can detect compact lenses more efficiently than single-lens analysis for $M_l\gtrsim 10^2-10^3M_\odot$ (Fig.~\ref{fig:forecast}). Although the CMB is a more powerful test of primordial black holes, LSD probes compact dark-matter halos and objects formed after recombination. It complements existing methods~\cite{Croon:2020wpr,Croon:2024rmw,Graham:2024hah} and can target specific theories, such as compact axion structures~\cite{Arvanitaki:2019rax}. 
Next-generation ground-based GW interferometers can improve current lensing limits by 5 orders of magnitude, potentially reaching the expected abundance of SMBHs. 
Strong limits can be expected for space-borne detectors~\cite{LISA:2017pwj,LISACosmologyWorkingGroup:2022jok} due to the high SNR and redshift of massive black hole mergers.

\new{LSD can be sought through a targeted matched-filter search, analogous to searches for sub-threshold counterparts of strongly-lensed signals~\cite{McIsaac:2019use,Dai:2020tpj,Li:2023zdl,LIGOScientific:2023bwz}. Rather than identifying individual counterparts, the goal is to detect an excess of matched-filter power --- with waveform fixed by the primary signal --- in a time window following each event, e.g.~by comparing the distribution of triggers before and after coalescence, accumulated over the catalogue.}
Several challenges must be addressed \new{in such a search}.
Uncertainties in the main signal (i.e.~parameter estimation, differences in the lens model) will degrade the SNR by the mismatch between the true and assumed waveforms.
Astrophysical GW backgrounds provide a source of confusion, as a small fraction of unresolved events can mimic LSD (i.e.~intrinsic parameters and sky localization very close to the main signal's, similar to the issue of false-alarm probability in strongly-lensed GWs~\cite{Wierda:2021upe,Caliskan:2022wbh}). 
LSD can still be distinguished by exploiting the correlation with the primary signal, absent in astrophysical backgrounds. 
\new{Alternatively, adapting} stochastic GW analysis techniques that leverage waveform information~\cite{Zhou:2022nmt,Zhou:2022otw,Zhong:2022ylh,Dey:2023oui} \new{to target power correlated with resolved events} will further alleviate the impact of detector glitches and astrophysical contamination.

LSD has interesting connections to strong lensing and microlensing.
The time distribution of secondary images (whether they appear after, around, or before the primary signal) could be used to reveal the image type~\cite{Lewis:2020asm,Williams_grb_microlensing_97}.
\new{Sources in dense environments, such as stellar-mass binaries in the discs of active galactic nuclei (AGN), could similarly probe the central massive black hole through repeated images and microlensing of their signals~\cite{Kocsis:2012ut,Gondan:2021fpr,Oancea:2022szu,Leong:2024nnx,Ubach:2025dob}.} 
\new{The framework carries over beyond geometric optics: in the time
domain, lensing convolves the signal with the Green's function of the
lens ensemble, in which each lens imprints a localized wave-optics lens
feature (WOLF) --- the smooth counterpart of a geometric image --- whose
delays and amplitudes follow the same statistics as the images
considered here (End Matter, Appendix E).}
Most interestingly, extending the analysis \new{to the wave-optics
regime} will provide insights into lighter lenses, including signatures of stars in strongly-lensed GWs~\cite{Diego:2019lcd,Mishra:2021xzz,Meena:2022unp,Cheung:2020okf,Yeung:2021roe,Shan:2022xfx,Shan:2023ngi,Shan:2023qvd,Meena:2023qdq,Zumalacarregui:2026uqs,LIGOScientific:2025rsn,Goyal:2025eqo,Shan:2025dcd,Chan:2025kyu,Hu:2025lhv,Chakraborty:2025pxt,Cheung:2026pky}.
\new{Recent studies have characterized diffractive imprints of primordial black holes and dark-matter (sub)halos on GW signals~\cite{Diego:2019rzc,Oguri:2020ldf,Oguri:2022zpn,Urrutia:2024pos,Ando:2026poq,Ando:2026eam,Guo:2022dre,Caliskan:2023zqm,Liu:2025ixi,Singh:2025uvp}, including the extension of stochastic diffraction to $10$--$10^4 M_\odot$ halos in the LISA band~\cite{Choi:2026lsa}.}

\begin{acknowledgements}
I am very grateful to S. Goyal, S. Savastano, H. Villarrubia-Rojo and Y. Wang for comments on the manuscript, to G. Brando, M. Cheung, J. Gair, G. Tambalo and M. Zaldarriaga for discussions, Han Gil Choi also for sharing the results of Ref.~\cite{GilChoi:2023qrz} \new{and the anonymous referees for valuable feedback}. This work relied on
\href{https://github.com/CosmoStatGW/gwfast}{\texttt{Gwfast}}~\cite{Iacovelli:2022bbs,Iacovelli:2022mbg}, \new{\href{https://github.com/miguelzuma/GLoW_public}{\texttt{GLoW}}~\cite{Villarrubia-Rojo:2024xcj},} \href{https://github.com/bradkav/PBHbounds}{\texttt{PBHbounds}}~\cite{kavanagh2019pbhbounds}, Astropy~\citep{astropy:2013, astropy:2018, astropy:2022}, Numpy~\cite{harris2020array} and Scipy~\cite{2020SciPy-NMeth}.
\new{The notebooks reproducing all results are publicly available at \url{https://github.com/miguelzuma/lsd-gw-notebooks}.}
\end{acknowledgements}

\appendix

\section*{End Matter}

\begin{figure}
    \centering
    \includegraphics[width=\columnwidth]{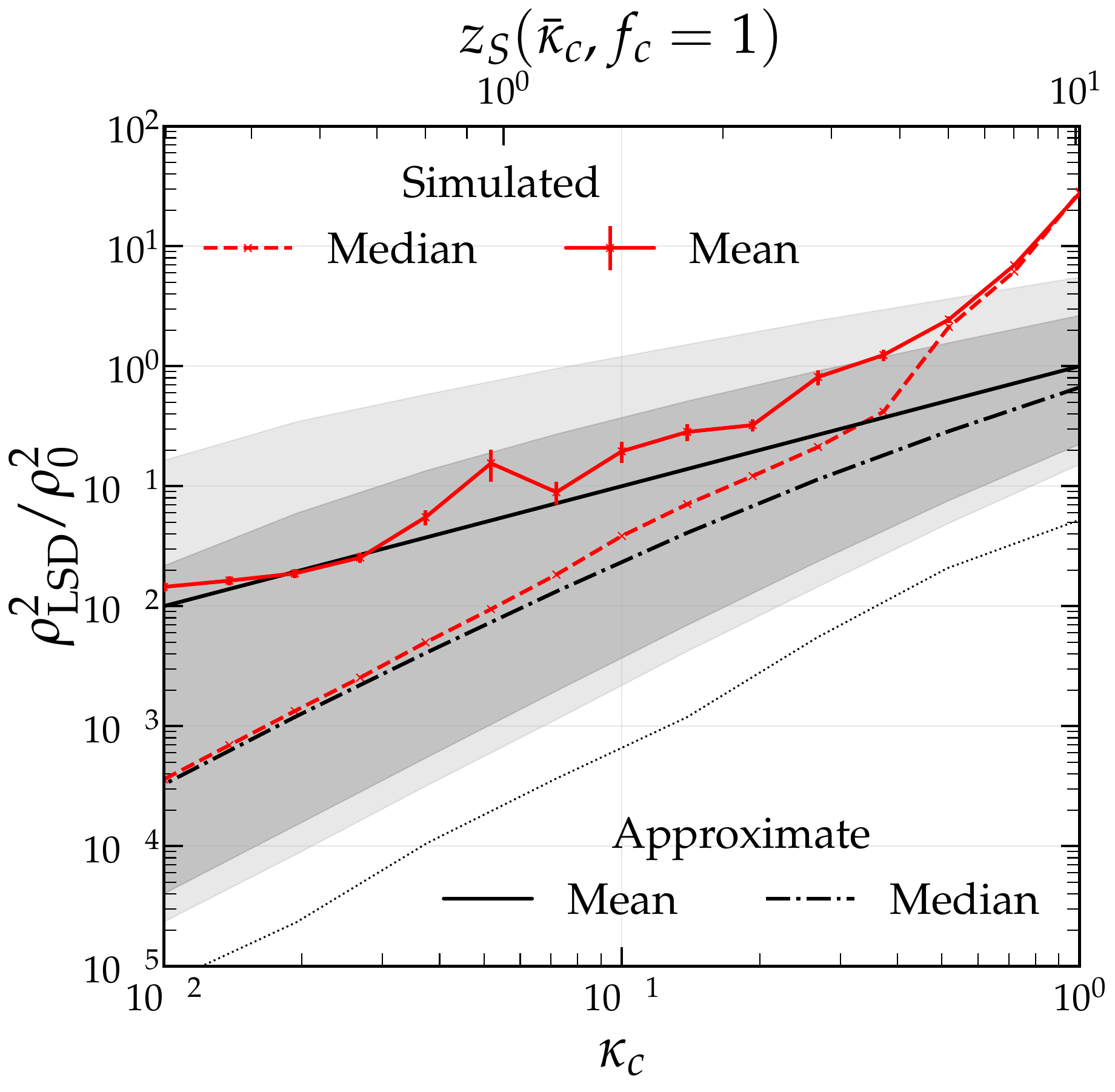}
    \caption{LSD-SNR distribution as a function of convergence.
    The mean and median of simulations (red) are larger than the isolated lens approximation (black) for $\kappa_c\gtrsim 0.1$.
    Error bars show the sampling uncertainty of the simulated mean, dominated by rare, closely aligned lenses.
    The 90\%, 98\% c.l. and minimum values of the approximate distribution are also shown (resp. gray bands and dotted line).
    The top scale shows the redshift at which $\bar\kappa_c=\kappa_c$ for $f_c=1$, Eq.~\eqref{eq:aconvergence_avg}.
    \notebooklink{https://nbviewer.org/github/miguelzuma/lsd-gw-notebooks/blob/main/LSD_realizations.ipynb\#Simulated-vs-approximate-LSD-power}
    }
    \label{fig:N0_SNR}
\end{figure}

\paragraph*{\bf A: statistics of the LSD-SNR.}
The realizations of Figs.~\ref{fig:image_realizations}~\&~\ref{fig:N0_SNR} draw a Poisson-distributed number of lenses with mean $\kappa_c\, y_{\rm max}^2$, distributed uniformly on the lens plane ($dN\propto y\,dy$) up to a maximum offset $y_{\rm max}$; the neglected contribution of more distant lenses scales as $y_{\rm max}^{-2}$ ($|\mu_-|\approx y^{-4}$) and is negligible for the $10$--$150$ lenses per realization used here.
In Fig.~\ref{fig:image_realizations} the images are computed exactly, as the stationary points of the joint Fermat potential of all lenses (projected onto a single plane, with magnifications given by its Hessian), using \texttt{GLoW}~\cite{Villarrubia-Rojo:2024xcj}; in Fig.~\ref{fig:N0_SNR} each lens contributes as if isolated (first sum of Eq.~\ref{eq:lsd_snr}), with $500$--$5000$ random realizations at each value of $\kappa_c$.
The agreement between both methods at low $\kappa_c$, and the brighter images formed at high $\kappa_c$, underpin the accuracy and conservativeness of the isolated-lens approximation used in the main text.

For $\mathcal{W}\to 1$, the integral in Eq.~\eqref{eq:lsd_rho2_avg} is exactly unity, $\int_0^\infty dy^2\,|\mu_-|(y) = 1$: the average relative LSD power equals the convergence, $\tilde\rho^2_{\rm LSD} = \bar\kappa_c$.
The distribution around this mean is highly skewed at low $\kappa_c$.
It is dominated by the closest lens, whose squared offset is exponentially distributed, $P(y_1^2 > s) = e^{-\kappa_c s}$, with median $y_1^2 = \ln 2/\kappa_c \gg 1$.
The median LSD power is thus $\tilde\rho^2_{\rm LSD} \approx |\mu_-(y_1)| \approx y_1^{-4} = (\kappa_c/\ln 2)^2 \propto \kappa_c^2$, far below the mean: typical realizations contain no closely aligned lens, and the average is set by rare, high-magnification configurations ($y_l\lesssim 1$).
The same configurations dominate the variance: from $|\mu_-|\to 1/(2y)$ at small offsets, the tail of the distribution is $P(\tilde\rho^2_{\rm LSD}>x) \approx \kappa_c/(4x^2)$, so the second moment diverges logarithmically and the sample mean converges slowly.
Cutting the tail at the largest value sampled in $N_{\rm r}$ realizations, $x_{\rm max}\simeq\sqrt{\kappa_c N_{\rm r}}/2$, the relative uncertainty of the mean decreases only as $\sim\sqrt{\ln(\kappa_c N_{\rm r})/(\kappa_c N_{\rm r})}$ ($28\%$ for $\kappa_c=10^{-2}$, $N_{\rm r}=5000$, in agreement with the $31\%$ scatter of independent sets of realizations).
The error bars of Fig.~\ref{fig:N0_SNR} are bootstrap estimates, which underestimate this scatter by a factor ${\sim}2$: resampling cannot produce alignments closer than the closest one realized.
The quantiles of the distribution are free of this limitation and converge rapidly.
In the opposite regime, $\kappa_c\gtrsim 0.1$, overlapping deflections increase the magnifications and form additional images~\cite{Katz_random_microlensing_86,Venumadhav:2017pps,Pascale:2021bdo}: the simulated LSD power exceeds the isolated-lens value (Fig.~\ref{fig:N0_SNR}), so treating lenses as isolated is conservative.

\paragraph*{\bf B: sensitivity computation.}
The LSD curves of Fig.~\ref{fig:forecast} are obtained from Eq.~\eqref{eq:rho_LSD} with $\bar\kappa_c(z) = f_c\,\tau(z)$, where $\tau(z)$ is the point-lens optical depth, Eq.~\eqref{eq:aconvergence_avg} at $f_c=1$.
Optimal SNRs and $P_{\rm det}$ are computed with \texttt{gwfast}~\cite{Iacovelli:2022bbs,Iacovelli:2022mbg} for each network and averaged over the ``power law + peak'' population \cite{Talbot:2018cva}, with $\rho_{\rm th}=8$.
The lens mass enters through the observable window: the fraction of the LSD power arriving after $T_{\rm min}$, as a function of the redshifted lens mass $M_{lz}\equiv (1+z_l)M_l$,
\begin{equation}\label{eq:feff}
f_{\rm eff}(M_{lz}) = \tilde\rho^{-2}_{\rm LSD}
\int dt\,\mathcal{W}(t)\,\frac{d\tilde\rho^2_{\rm LSD}}{dt}\,,
\end{equation}
is evaluated from Eq.~\eqref{eq:lsd_snr_dt} at $z_S=1$ (below the typical source redshift of future detectors, which pushes delays to larger values: a conservative choice).
Since secondary signals are Poisson-distributed with rate $\propto f_c$, a null observation over a time $T_{\rm obs}$ excludes, at 90\% c.l.,
\begin{equation}\label{eq:fc90}
f_c > \frac{2.3}{\rho^2_{\rm LSD,pop}(f_c{=}1)\,
f_{\rm eff}(M_{lz})\, T_{\rm obs}}\,.
\end{equation}
The dependence of this limit on the observable window, on the redshift evolution of the merger rate and on the lens mass function is quantified in turn below.

\begin{figure}[t]
    \centering
    \includegraphics[width=\columnwidth]{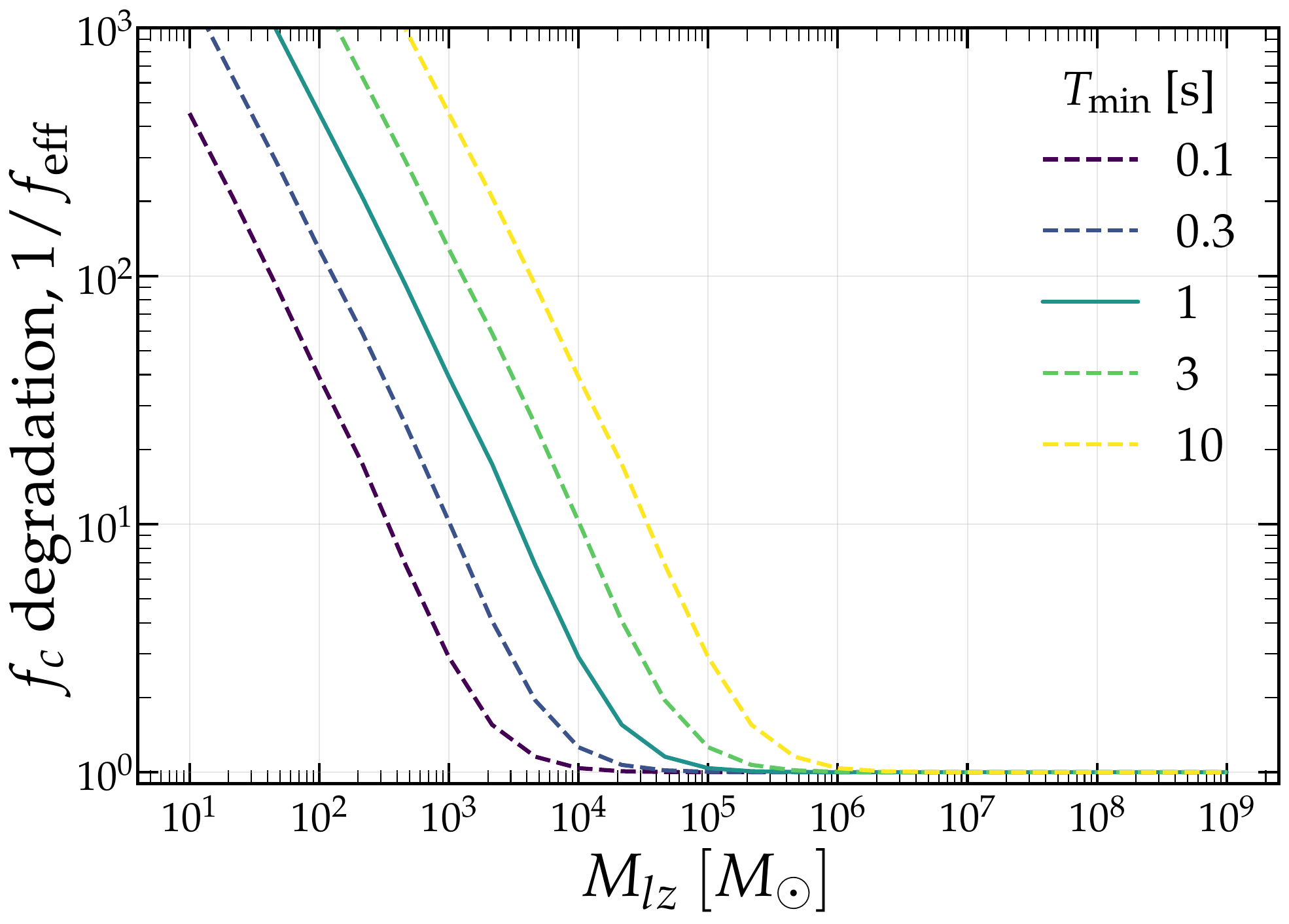}
    \caption{Degradation of the $f_c$ sensitivity due to the observable window (Eq.~\ref{eq:feff}, $z_S=1$), for different choices of $T_{\rm min}$.
    The fiducial value (1\,s, solid) affects only $M_{lz}\lesssim 10^5 M_\odot$; varying $T_{\rm min}$ by an order of magnitude shifts the low-mass edge proportionally, leaving the high-mass limits unchanged.
    \notebooklink{https://nbviewer.org/github/miguelzuma/lsd-gw-notebooks/blob/main/LSD_forecast.ipynb\#Sensitivity-to-the-time-window-Tmin}
    }
    \label{fig:tmin}
\end{figure}

\textit{Observable window}: Figure~\ref{fig:tmin} shows the degradation $1/f_{\rm eff}$ for window choices $T_{\rm min}\in[0.1,10]$\,s.
The sensitivity is unaffected for $M_{lz}\gtrsim 10^5\,M_\odot\,(T_{\rm min}/1\,{\rm s})$; below, the window removes the typical delays (Eq.~\ref{eq:time_delay}), and changing $T_{\rm min}$ only shifts the low-mass edge of Fig.~\ref{fig:forecast} in proportion.
An upper cutoff $T_{\rm max}$ (set, e.g., by the duration of the observing run or by the validity of the constant antenna-pattern approximation) analogously suppresses the sensitivity to high lens masses, $M_{lz}\gtrsim 10^{7}\,(T_{\rm max}/1\,{\rm d})\,M_\odot$.

\textit{Merger-rate evolution}: Table~\ref{tab:rates} quantifies the dependence on the assumed redshift evolution of the merger rate, $f_\text{merge}(z)$ in Eq.~\eqref{eq:rho_LSD}, in the large-mass limit where the window is irrelevant ($f_{\rm eff}\to 1$).
The fiducial constant comoving rate gives the weakest limits for every network and is therefore adopted throughout.
A star-formation-rate evolution~\cite{Madau:2014bja}, $f_\text{merge}\propto (1+z)^{2.7}/[1+((1+z)/2.9)^{5.6}]$ with no delay between formation and merger, strengthens them by a factor ${\sim}4$, reflecting the higher merger rate at the redshifts that dominate the detected population.
A pure power law $(1+z)^{2.7}$, which matches the star-formation rate at low redshift but never turns over, is nearly degenerate with it for LVK-O5 (whose horizon lies below the turnover) but strengthens the 3G limits by ${\sim}400$: the high-redshift behaviour of the merger rate, which current observations do not constrain~\cite{KAGRA:2021duu}, is the dominant astrophysical uncertainty for ET and CE.

\begin{figure}[t]
    \centering
    \includegraphics[width=\columnwidth]{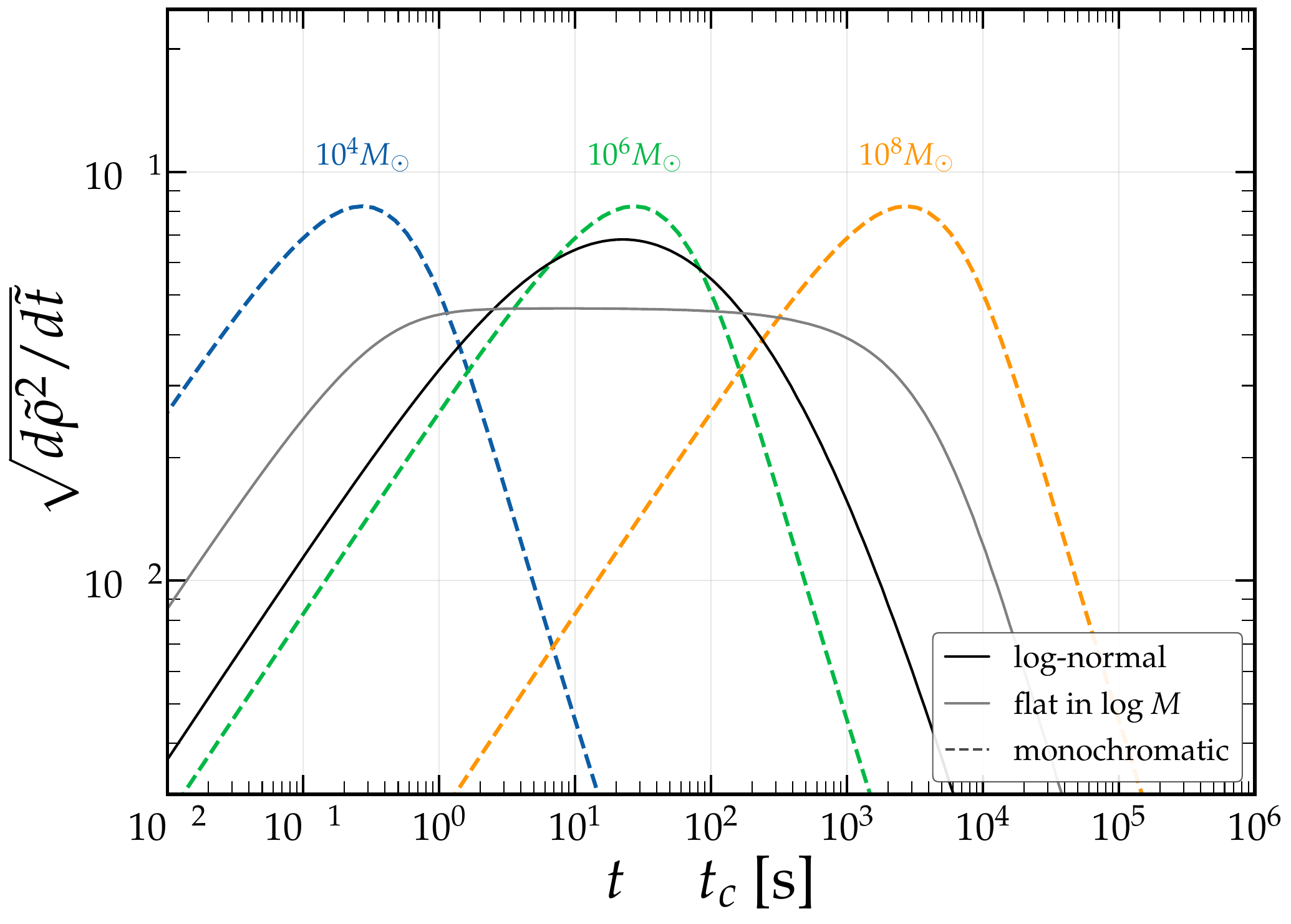}
    \caption{Temporal distribution of the LSD power for extended lens mass functions ($z_S=1$, $f_c=1$): a log-normal distribution ($\sigma=0.5$ dex around $10^6 M_\odot$, black) and a flat-in-the-logarithm spectrum ($df_c/d\log M_{lz}={\rm const}$ over $10^4$--$10^8 M_\odot$, gray), both with the same total abundance.
    Dashed lines show monochromatic populations (as in Fig.~\ref{fig:lsd_pedagogical}), whose superposition gives the extended-spectrum predictions.
    \notebooklink{https://nbviewer.org/github/miguelzuma/lsd-gw-notebooks/blob/main/LSD_exploration.ipynb\#Extended-lens-mass-functions}
    }
    \label{fig:massfunction}
\end{figure}

\textit{Lens mass function}: because Eq.~\eqref{eq:lsd_snr_dt} is linear in the lens mass function, predictions for extended mass spectra follow by superposing monochromatic populations.
Figure~\ref{fig:massfunction} shows the temporal distribution of the LSD power for two extended mass functions, normalized to the same total abundance $f_c$: a log-normal distribution of width $\sigma = 0.5$ dex centered on $M_{lz} = 10^6\,M_\odot$, and a flat-in-the-logarithm spectrum, $df_c/d\log M_{lz} = {\rm const}$, both spanning $M_{lz}\in [10^4, 10^8]\,M_\odot$.
Each mass scale enters at its characteristic delay, $\tilde t_l \propto M_{lz}$ (Eq.~\ref{eq:time_delay}): the log-normal signal closely resembles that of its central mass, while the flat spectrum spreads the same LSD power evenly along the envelope of the monochromatic components.
The temporal distribution of an observed LSD signal thus encodes the lens mass spectrum.

\begin{table}[b]
\caption{\label{tab:rates}Sensitivity to the merger-rate evolution: 90\% c.l.\ limits on $f_c$ after 1 yr, in the large-mass limit ($f_{\rm eff}\to1$, i.e.~the plateaus of Fig.~\ref{fig:forecast}), for a constant comoving rate (fiducial), the star-formation rate~\cite{Madau:2014bja} and a pure power law $(1+z)^{2.7}$.
All rates are normalized to the same local value $\mathcal{R}_0$.}
\begin{ruledtabular}
\begin{tabular}{lccc}
 & $f_\text{merge}=1$ & SFR & $(1+z)^{2.7}$ \\
\hline
LVK-O5 & $1.1\times10^{-3}$ & $2.7\times10^{-4}$ & $2.5\times10^{-4}$ \\
ET & $1.4\times10^{-6}$ & $3.2\times10^{-7}$ & $3.4\times10^{-9}$ \\
CE & $6.1\times10^{-7}$ & $1.5\times10^{-7}$ & $1.3\times10^{-9}$ \\
\end{tabular}
\end{ruledtabular}
\end{table}

\paragraph*{\bf C: comparison with single-image searches.}
The brightest secondary image is individually detectable when $\sqrt{|\mu_-|}\,\rho_0 > \rho_{\rm img}$.
In terms of the observed (magnified) SNR of the primary, $\rho_{\rm obs} = \sqrt{\mu_+}\,\rho_0$, the condition reads $|\mu_-/\mu_+| > (\rho_{\rm img}/\rho_{\rm obs})^2$ and defines a critical offset $y^2_{\rm cr}(\rho_{\rm obs})$ below which a lens produces a detectable counterpart: the expected number of counterparts per event is $\bar\kappa_c\, y^2_{\rm cr}$ (Eq.~\ref{eq:enclosed_N}).
The rate of detectable counterparts is the analog of Eq.~\eqref{eq:rho_LSD}, with the windowed LSD power replaced by the expected number of counterparts,
\begin{align}\label{eq:Nimg}
\dot N_{\rm img} &= \mathcal{R}_0 f_c \int d\vec \theta\,
\frac{dP}{d\vec\theta}\int dz\, \tau\,\frac{dV}{dz}\, f_\text{merge}\;
\overline{y^2_{\rm cr}}\,,
\\ \label{eq:ycr_avg}
\overline{y^2_{\rm cr}} &= \int dw\, \frac{dP}{dw}\;
\Theta\!\left(w\,\rho_{\rm opt}-\rho_{\rm th}\right)\,
y^2_{\rm cr}\!\left(w\,\rho_{\rm opt}\right)\,,
\end{align}
where $\tau(z) = \bar\kappa_c(z)/f_c$ is the optical depth, $w$ relates the optimal and observed SNRs ($\rho_{\rm obs} = w\,\rho_{\rm opt}$, distributed as $dP/dw$ over sky position, inclination and polarization), $\Theta$ requires the detection of the primary event, and the critical offset is restricted to the observable window, $T_{\rm min} < t_l(y) < T_{\rm max}$ (Eq.~\ref{eq:time_delay}).
Counterparts are Poisson-distributed with rate $\dot N_{\rm img}\propto f_c$; a null search excludes $f_c > 2.3/[\dot N_{\rm img}(f_c{=}1)\,T_{\rm obs}]$ at 90\% c.l.

\begin{figure}[t]
    \centering
    \includegraphics[width=\columnwidth]{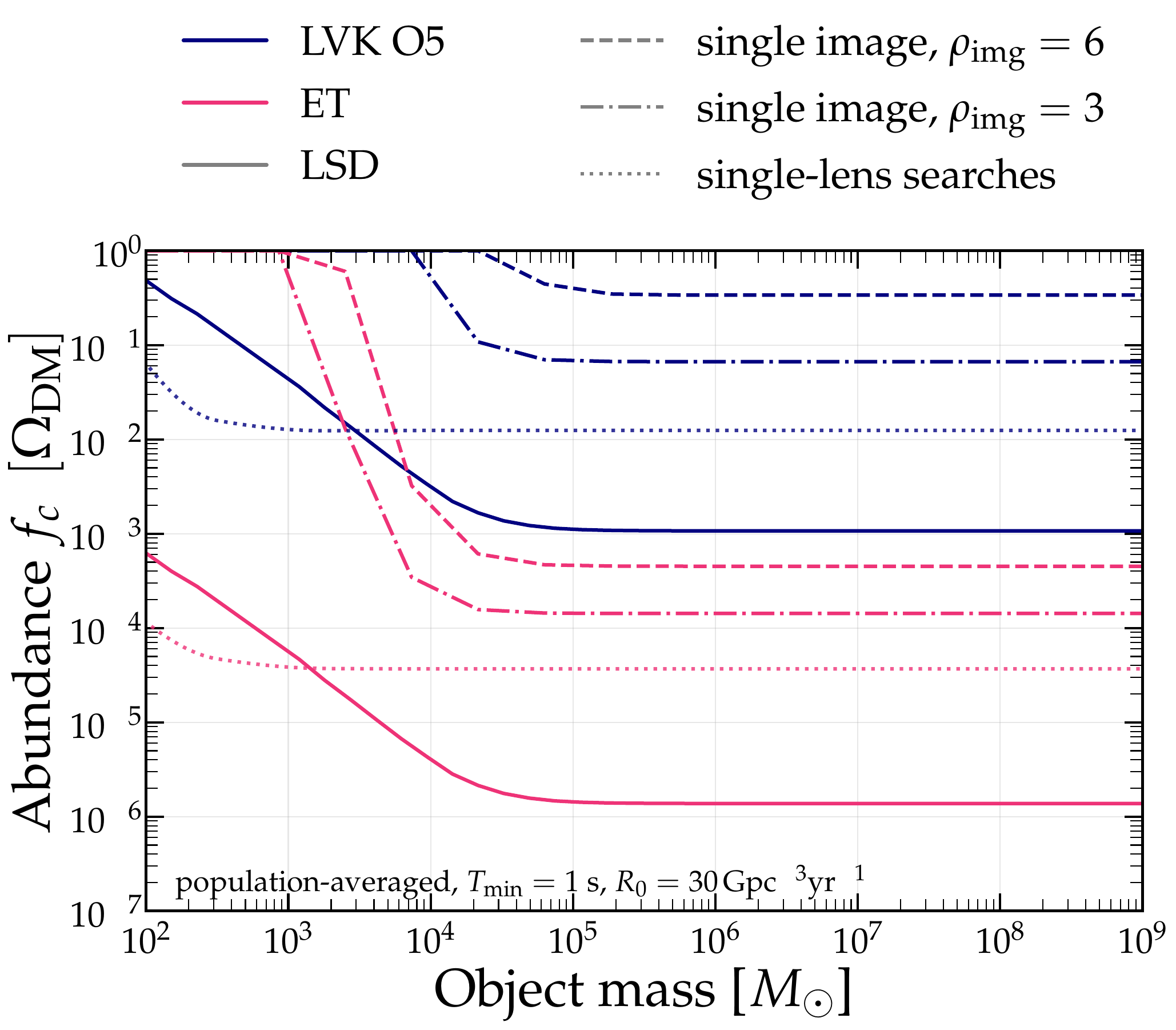}
    \caption{Sensitivity of LSD (solid) and of counting individually detectable secondary images, computed with identical population averages and window ($T_{\rm min}=1$\,s), with $\rho_{\rm th}=8$ and counterpart thresholds $\rho_{\rm img}=6$ (dashed) and $3$ (dash-dotted); 90\% c.l.\ limits for 1 year of observation, axes as in Fig.~\ref{fig:forecast}.
    Dotted lines show dedicated single-lens forecasts~\cite{Jung:2017flg,GilChoi:2023qrz} (LVK-O5 as in Fig.~\ref{fig:forecast}, $\ln\Lambda=3$ for ET).
    CE (not shown) is qualitatively similar to ET.
    \notebooklink{https://nbviewer.org/github/miguelzuma/lsd-gw-notebooks/blob/main/LSD_forecast.ipynb\#Comparison-with-single-image-searches}
    }
    \label{fig:single_image}
\end{figure}

Figure~\ref{fig:single_image} compares both statistics under identical assumptions ($\rho_{\rm th}=8$, $\rho_{\rm img}=6$, $T_{\rm min}=1$\,s, same population averages and networks; CE, qualitatively similar to ET, is omitted for clarity).
The counterpart threshold is set below the detection threshold because the search is targeted: the waveform and arrival window are constrained by the primary event, reducing the false-alarm rate.
LSD improves over image counting by ${\sim}2.5$ orders of magnitude for all three networks (a factor ${\sim}320$--$340$ at $\rho_{\rm img}=6$).
The counting limits flatten at high mass like the dedicated single-lens forecasts~\cite{Jung:2017flg,GilChoi:2023qrz} (dotted).
In that regime $y^2_{\rm cr}\propto \rho_{\rm obs}/\rho_{\rm img}$ (from $|\mu_-|\approx y^{-4}$), so lowering the threshold strengthens the counting limits: for $\rho_{\rm img}=3$ (dash-dotted) they land within a factor ${\sim}4$--$5$ of the dedicated forecasts, which further employ sub-threshold statistics ($\ln\Lambda = 3$ in Refs.~\cite{Jung:2017flg,GilChoi:2023qrz}) and magnification bias in the event selection.
Even against this aggressive counting estimate, LSD retains a gain of ${\sim}2$ orders of magnitude (a factor $60$--$120$).
The LSD gain comes from accumulating the power of many faint images over every detected event, rather than waiting for rare, individually bright counterparts: the information is gathered from the whole catalog, not from the small subset of events with a detectable counterpart.

\begin{figure}[t!]
    \centering
    \includegraphics[width=\columnwidth]{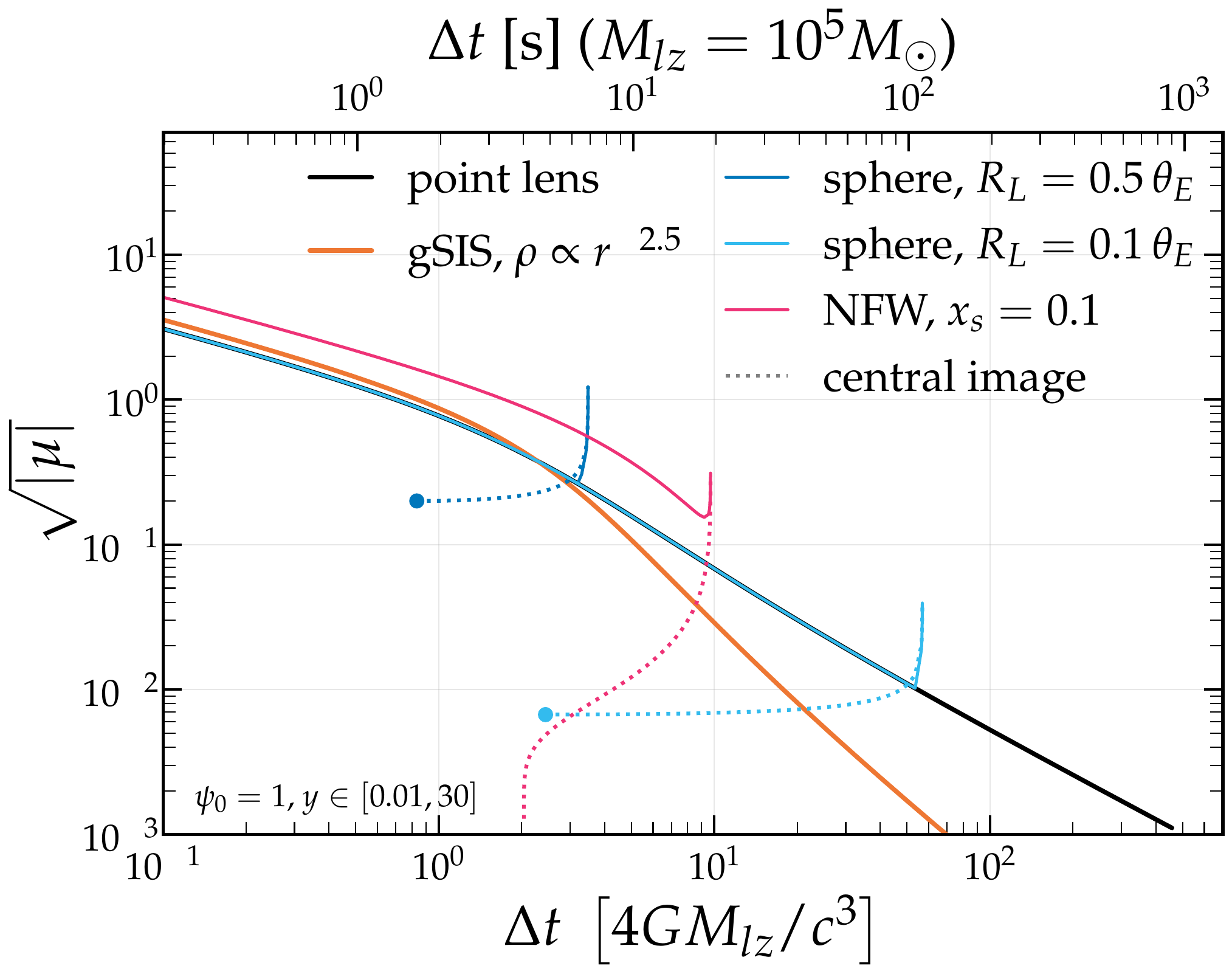}
    \caption{Additional images of extended lenses: amplitude vs time delay along tracks of varying impact parameter, $y\in[0.01,30]$, for lenses with a common Einstein radius ($\psi_0=1$).
    Solid/dotted lines show saddle-point/central images; spikes mark radial caustics, beyond which no additional images form; circles mark the perfect-alignment limit ($y\to 0$) of the spheres' central images, which retain a finite magnification; the cuspy NFW central image is instead completely demagnified in this limit~\cite{Tambalo:2022wlm}.
    The top axis shows physical delays for $M_{lz}=10^5 M_\odot$.
    \notebooklink{https://nbviewer.org/github/miguelzuma/lsd-gw-notebooks/blob/main/LSD_generalizations.ipynb\#Extended-lenses}
    }
    \label{fig:extended}
\end{figure}

\paragraph*{\bf D: extended lenses.}
Figure~\ref{fig:extended} shows the amplitude and time delay of the additional images of extended lenses, as tracks of varying impact parameter $y\in[0.01,30]$, computed with \texttt{GLoW}~\cite{Villarrubia-Rojo:2024xcj}.
Four profiles are compared to the point lens: a generalized singular isothermal sphere (gSIS) with $\rho\propto r^{-5/2}$, i.e.~deflection $\alpha(x) = \psi_0\,x^{-1/2}$~\cite{Tambalo:2022wlm}, homogeneous spheres (constant 3D density) of radius $R_L = 0.5$ and $0.1\,\theta_E$, and a Navarro--Frenk--White (NFW) profile~\cite{Navarro:1996gj} with scale radius $x_s = 0.1\,\theta_E$~\cite{Brando:2024inp}.
All lenses share the same Einstein radius ($\psi_0 = 1$, equal projected mass within $\theta_E$): populations with equal number density then have equal optical depths and identical offset statistics, so the density profile affects the LSD signal \textit{only} through the magnification--delay relations shown.
Note that offsets are Poisson-distributed, so the density of lenses at a given point along a track grows $\propto y$ (Eq.~\ref{eq:enclosed_N}), weighting each population toward the large-delay end of its track.
Equal number density does, however, correspond to different cosmic mass densities: the total (virial) mass of an extended lens can greatly exceed the projected mass within its Einstein radius.
Normalizing populations to equal mass density instead rescales the number of lenses --- lowering $\bar\kappa_c$ and hence the amplitude of $d\tilde\rho^2_{\rm LSD}/dt$ --- without affecting the magnification--delay relations shown.
The resulting degradation of the abundance sensitivity is given for each profile below.

For an axisymmetric lens the inverse magnification factorizes as
\begin{equation}\label{eq:mu_eigenvalues}
\mu^{-1} = \left(1-\frac{\psi'}{x}\right)\left(1-\psi''\right)\,,
\end{equation}
where the first (tangential) eigenvalue describes image distortions along the Einstein ring and vanishes on it, and the second (radial) eigenvalue describes distortions perpendicular to the ring.
For $y\ll 1$ all additional images lie near the Einstein ring: the delay is dominated by the geometric term, $\Delta t \approx 2y$ for \textit{any} profile, the tangential factor of the saddle is $\approx y$, and the profile enters only through the radial factor $1-\psi''(1)$.
As $y$ grows the saddle moves toward the lens center, probing the inner density profile; cored profiles develop a radial caustic ($\psi''=1$), where the saddle and central images merge and annihilate.

\textit{Point lens} ($\psi=\ln x$): the radial factor is $1-\psi''(1) = 2$; at large offsets $|\mu_-|\approx y^{-4}$ while $\Delta t\approx y^2/2$, so the track falls as $\sqrt{|\mu|}\propto \Delta t^{-1}$; the singular density forms neither a central image nor a radial caustic.
All the mass lies within the Einstein radius, so the limits of Fig.~\ref{fig:forecast} apply as stated.
The \textit{generalized singular isothermal sphere} ($\psi = 2\sqrt{x}$) is qualitatively similar, as the central cusp also prevents a radial caustic.
The shallower profile gives a radial factor $3/2$ ($\kappa(\theta_E)=1/4$), making the saddle brighter by exactly $4/3$ at fixed, small $\Delta t$.
At large offsets the trend reverses: the saddle approaches the cusp, $|x_-|\approx y^{-2}$ (from $\alpha(x_-) = |x_-| + y$), both eigenvalues grow as $y^{3}$ and $|\mu|\simeq 2\,y^{-6}$, so $\sqrt{|\mu|}\propto \Delta t^{-3/2}$ and the tracks cross at $\Delta t \approx 8GM_{lz}$.
The enclosed mass of the cusp grows as $M(<r)\propto r^{1/2}$, so for a profile truncated at $R_t$ only a fraction $\sim(R_E/R_t)^{1/2}$ of the total mass forms images, where $R_E = \theta_E D_L$ is the Einstein radius: at fixed $f_c$ the limits weaken by this mild factor.

\textit{Homogeneous spheres} deform the point-lens track: while the saddle lies outside the sphere ($|x_-|\gtrsim R_L$) the tracks coincide, but once it enters the sphere the surface density lowers the radial eigenvalue, brightening the image up to the radial caustic (bright spikes at $\Delta t\approx 3$ and $\approx 50$ for $R_L=0.5$ and $0.1\,\theta_E$ in Fig.~\ref{fig:extended}).
The caustic lies at $y_r\approx \theta_E/R_L$, cutting the LSD power to $99\%$, $91\%$ and $77\%$ of the point-lens value for $R_L = 0.1$, $0.3$ and $0.5\,\theta_E$, and to zero for $R_L\gtrsim\theta_E$.
The corresponding delay, $\Delta t_r \approx 2GM_{lz}\,(\theta_E/R_L)^{2}$ (accurate for $R_L\lesssim 0.3\,\theta_E$), must also exceed the window: requiring $\Delta t_r > T_{\rm min}$ makes the constraining power mutually dependent on the lens' mass and size, $M_{lz}\gtrsim 10^{5}\,(R_L/\theta_E)^{2}\,(T_{\rm min}/1\,{\rm s})\,M_\odot$.
A similar cutoff exists for \textit{Navarro--Frenk--White}: the extended, shallow density profile brightens the saddle along the whole track, which ends at the radial caustic ($\Delta t\approx 10$ for $x_s=0.1\,\theta_E$).
Additional images require a near-critical central column density, $\Sigma \sim \Sigma_{\rm cr} = [4\pi G\, d_{\rm eff}]^{-1} \approx 0.6\,{\rm g\,cm^{-2}}$ for sources at $z_S=1$, with $d_{\rm eff} = D_L D_{LS}/D_S$: the multiple-image cross section of NFW halos falls orders of magnitude below that of isothermal profiles and is extremely sensitive to the central density~\cite{Li:2000dw}.
Beyond the radial caustic cored lenses form no additional images, cutting the LSD signal at the corresponding delay.
Central images (dotted) are strongly demagnified and arrive at short delays, where wave-optics corrections may be relevant~\cite{Tambalo:2022wlm}.

Geometric-optics LSD therefore probes only very concentrated halos; more diffuse halos imprint wave-optics lens features instead (Appendix~E).
The point lens has the flattest, most detectable large-delay tail, making it the optimistic reference for Fig.~\ref{fig:forecast}; conversely, a cutoff or steepening in the temporal distribution of an observed LSD signal would measure the size and profile of the lenses.

\paragraph*{\bf E: generalization to wave-optics.}
\begin{figure*}[t]
    \centering
    \includegraphics[width=\textwidth]{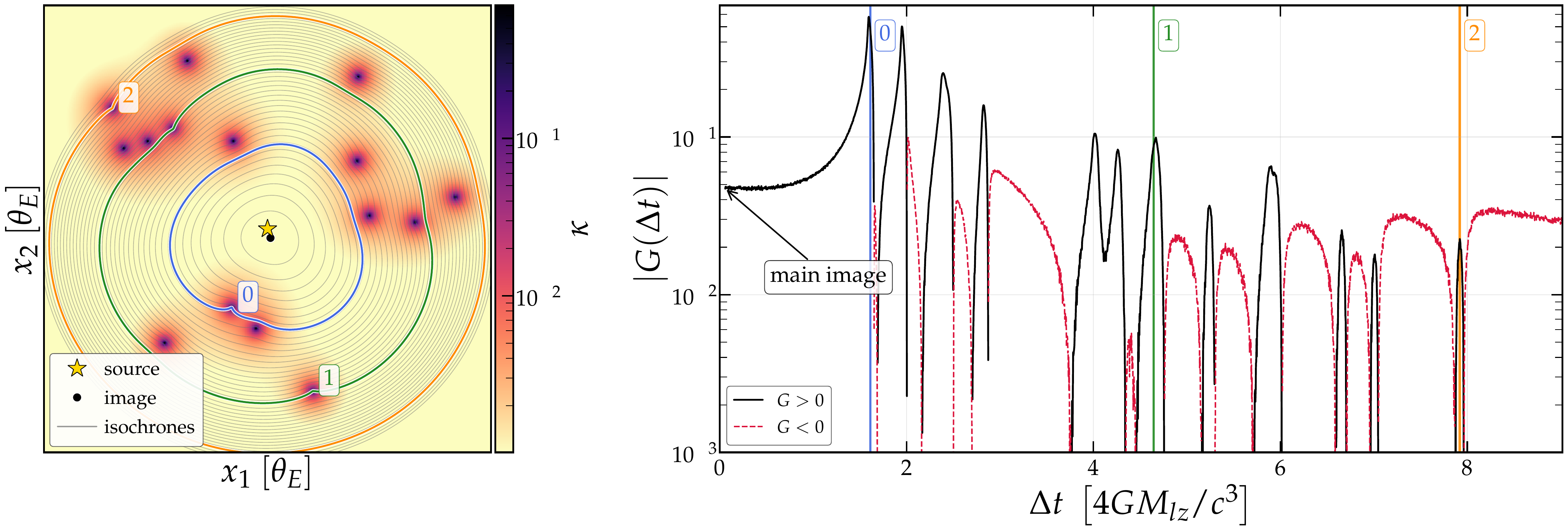}
    \caption{Wave-optics generalization of LSD.
    \textbf{Left:} projected density of a realization of 15 NFW subhalos ($r_{\rm vir}/r_s = 50$, equal masses, total $\psi_0=1$) surrounding the source (star, at the origin); gray lines show isochrones of the Fermat potential, the black circle the main image (a global minimum, the only image), and colored curves the isochrones through three subhalos (labeled by increasing delay).
    \textbf{Right:} the Green's function of the lens ensemble (Eq.~\ref{eq:greens}), with $G \to \sqrt{\mu}\,\delta(\Delta t)$ in the single-image geometric-optics limit; negative stretches of $G$ are shown dashed (red).
    Each subhalo imprints a wave-optics lens feature (WOLF) at the delay of the isochrone crossing it.
    \notebooklink{https://nbviewer.org/github/miguelzuma/lsd-gw-notebooks/blob/main/LSD_generalizations.ipynb\#Wave-optics-Greens-function-of-a-lens-ensemble}
    \label{fig:wolf}}
\end{figure*}

Beyond geometric optics, the lensed strain is the convolution of the unlensed signal with the Green's function of the lens ensemble,
\begin{equation}\label{eq:greens}
h_L(t) = \int d\Delta t\; G(\Delta t)\, h_0(t-\Delta t)\,,\quad
G \propto \frac{dI}{d\Delta t}\,,
\end{equation}
where $I(\Delta t)$ is the area enclosed by the isochrones of the Fermat potential~\cite{Tambalo:2022plm,Villarrubia-Rojo:2024xcj}.
In the geometric-optics limit the Green's function recovers Eq.~\ref{eq:lensed_time_domain}: minima and maxima contribute $G \propto \pm\sqrt{\mu_I}\,\delta(\Delta t - t_I)$, while a saddle point produces a divergence, $G \propto \sqrt{|\mu_I|}\,|\Delta t - t_I|^{-1}$, that distorts the image into the Hilbert-transformed waveform $h_0^\dagger$~\cite{Ulmer:1994ij,Tambalo:2022plm,Villarrubia-Rojo:2024xcj}.
A finite signal bandwidth smooths these singular features over ${\sim}(2\pi f)^{-1}$.
The imprints of extended cosmic structures on weakly lensed signals were studied in this framework in Ref.~\cite{Savastano:2023spl}, whose Fig.~5 first presented the correspondence between individual structures and localized features of $G$.

Figure~\ref{fig:wolf} shows a realization of 15 equal NFW subhalos with concentration $r_{\rm vir}/r_s = 50$ (total $\psi_0 = 1$) surrounding the source, computed with \texttt{GLoW}~\cite{Villarrubia-Rojo:2024xcj}.
The subhalos are subcritical, so no additional images form (the global minimum is the only image), yet each subhalo along the line of sight imprints a localized \textit{wave-optics lens feature} (WOLF) on $G$, at the delay of the isochrone crossing it --- the wave-optics counterpart of the faint additional images of the main text.

The LSD construction carries over to the wave-optics regime: lens positions remain Poisson-distributed, and the delays and amplitudes of WOLFs follow relations similar to those of the geometric images (Eqs.~\ref{eq:enclosed_N},~\ref{eq:lsd_snr_dt}), retaining the sensitivity to the abundance, mass and density profile of the lenses.
The delay--amplitude statistics and the population forecasts of the main text generalize accordingly; matched-filter detection (with $h_0^\dagger$ replaced by the band-limited WOLF imprint) remains possible for WOLFs that do not overlap the main signal.
This extends the LSD program to light lenses, $2\pi f\, \Delta t \lesssim 1$, where the geometric description breaks down and the windowed analysis of the main text loses sensitivity.

WOLFs can be modeled accurately by combining theoretical descriptions with numerical wave-optics computations, as demonstrated for stochastic diffraction by dark-matter halos~\cite{Choi:2026lsa} and stellar fields~\cite{Zumalacarregui:2026uqs}.

\bibliography{gw_lensing}

\end{document}